\documentclass[twocolumn,preprintnumbers,showpacs]{revtex4}
\usepackage{amsmath}
\usepackage{dcolumn}
\usepackage{bm}
\usepackage{epsfig}
\usepackage{epstopdf}
\usepackage{graphicx}
\usepackage{amsfonts}
\usepackage{amssymb}
\usepackage{mathrsfs}
\usepackage{color}
\usepackage{multirow}
\usepackage{appendix}

\begin{document}
\title{Experimental verification of a Jarzynski-related information-theoretic equality using a single trapped ion}
\author{T. P. Xiong$^{1,2}$}
\author{L. L. Yan$^{1}$}
\author{F. Zhou$^{1}$}
\email{zhoufei@wipm.ac.cn}
\author{K. Rehan$^{1,2}$}
\author{D. F. Liang$^{1,3}$}
\author{L. Chen$^{1}$}
\author{W. L. Yang$^{1}$}
\author{Z. H. Ma$^{4}$}
\author{M. Feng$^{1,3,5,6}$}
\email{mangfeng@wipm.ac.cn}
\author{V. Vedral$^{7,8,9}$}
\email{vlatko.vedral@qubit.org}
\affiliation{$^{1}$ State Key Laboratory of Magnetic Resonance and Atomic and Molecular Physics,
Wuhan Institute of Physics and Mathematics, Chinese Academy of Sciences, Wuhan, 430071, China\\
$^{2}$ School of Physics, University of the Chinese Academy of Sciences, Beijing 100049, China \\
$^{3}$ Synergetic Innovation Center for Quantum Effects and Applications (SICQEA), Hunan Normal University,
Changsha 410081, China \\
$^{4}$ Department of Mathematics, Shanghai Jiaotong University, Shanghai, 200240, China \\
$^{5}$ Center for Cold Atom Physics, Chinese Academy of Sciences, Wuhan 430071, China \\
$^{6}$ Department of Physics, Zhejiang Normal University, Jinhua 321004, China \\
$^{7}$ Department of Physics, Clarendon Laboratory, University of Oxford, Parks Road, Oxford OX1 3PU, United Kingdom \\
$^{8}$ Centre for Quantum Technologies, National University of Singapore, 117543, Singapore \\
$^{9}$ Department of Physics, National University of Singapore, 2 Science Drive 3, 117551, Singapore }

\begin{abstract}
Most non-equilibrium processes in thermodynamics are quantified only by inequalities, however the Jarzynski relation presents a remarkably simple and general equality relating non-equilibrium quantities with the equilibrium free energy, and this equality holds in both classical and quantum regimes. We report a single-spin test and confirmation of the Jarzynski relation in quantum regime using a single ultracold $^{40}Ca^{+}$ ion trapped in a harmonic potential, based on a general information-theoretic equality for a temporal evolution of the system sandwiched between two projective measurements. By considering both initially pure and mixed states, respectively, we verify, in an exact and fundamental fashion, the non-equilibrium quantum thermodynamics relevant to the mutual information and Jarzynski equality.
\end{abstract}
\maketitle

Since the original proposals of the celebrated ideas of Maxwell's demon \cite{demon} and Szil{\'a}rd's engine \cite{engine}, much effort has been devoted to incorporating information into thermodynamics by reconsidering the meaning of thermodynamical entropic and energetic costs. So far, the field of information thermodynamics has reformulated the restrictions of the original thermodynamics, e.g., the second law of thermodynamics, in the light of the interplay between the amount of information and its thermodynamical utility. Reconsideration of the second law based on the notion of information and further clarifications of the physical nature of information are typically expected to reconcile any apparent contradictions we might have regarding our understanding of the laws of thermodynamics \cite{parrondo}.

In addition to this, there has been a parallel line of development because the conventional equilibrium thermodynamics cannot reasonably treat most natural or engineered processes that occur far from equilibrium.  Namely, the non-equilibrium processes in thermodynamics are usually described by inequalities (or equalities that only hold in the linear regime, which means not far from equilibrium). In contrast, the Jarzynski relation presents a simple and general equality to calculate the free-energy difference between two states from Boltzmann-weighted statistics of the irreversible work done along the trajectories arbitrarily out of equilibrium \cite{jarzynski}. As the only equality in non-equilibrium thermodynamics, the Jarzynski relation can also be understood from the fluctuation theorem \cite{crooks} under the assumption of microscopically reversible and thermostated dynamics. The ensuing investigations \cite{exp1,exp2,exp3,exp4} have further confirmed that the Jarzynski equality promises to correctly predict any behavior, adiabatic or arbitrarily fast, in the presence of the Boltzmann statistics. A comprehensive review of thermodynamic experiments regarding the fluctuation theorem can be found in \cite{soft}.

Understanding thermodynamical process at the quantum level is currently a topic attracting much attention \cite{book,rev-1,rev-2}. Several attempts have been paid to extend the Jarzynski relation to quantum regime \cite{quant-th1,quant-th2,quant-3,quant-4,quant-5,quant-6,quant-exp}. From the quantum perspective, the origin of fluctuations is no
longer just thermal but also quantum, and most thermodynamical quantities should be retraced. For example, the amount of work itself is not an observable in quantum thermodynamics, and its quantification therefore needs to be reconsidered \cite{quant-4,quant-5,meas1,meas4}. Besides this, the quantum entropy is actually an indication of the entanglement between the system and its environment \cite{ent-1}, and is no longer simply the thermodynamical arrow of time.

Here we show a single-spin verification of an information-theoretic equality relevant to Jarzynski relation via experimental manipulation of a trapped-ion system. Ultracold trapped ions represent an ideal tool to investigate the thermodynamics \cite{quant-exp,meas1,ion-thermo-2,ion-thermo-3,ion-thermo-4}. In comparison with a previous attempt \cite{quant-exp} using both the spin and vibrational degrees of freedom of a trapped ion to explore the Jarzynski equality, our execution only on a qubit (i.e., a single spin) encoded in a single ultracold $^{40}Ca^{+}$ ion provides a more fundamental test of the information-theoretic equality (not just the Jarzynski equality itself) in a closed quantum system. This makes sure that our experimental verification of the Jarzynski-relevant equality is made under an ideal fluctuation theorem in the absence of decoherence. Consequently, our manipulation could demonstrate, in a perfectly quantum mechanical way, the interplay between non-equilibrium phenomena and information at the nanoscale \cite{review}.

Our work is mainly based on a previous proposal \cite{vedral} of a general quantum mechanical process involving a temporal evolution sandwiched by two projective measurements. Since the measurement updates the original state to a new state with information encoded \cite{parrondo}, this is a typical process of thermodynamics of information. In general, the updated state, after the measurement, is out of equilibrium even if the system is initially prepared in an equilibrium state. So the process of interest definitely reflects non-equilibrium thermodynamics.

\begin{figure}[t]
\centering {\includegraphics[width=8 cm, height=5.8 cm]{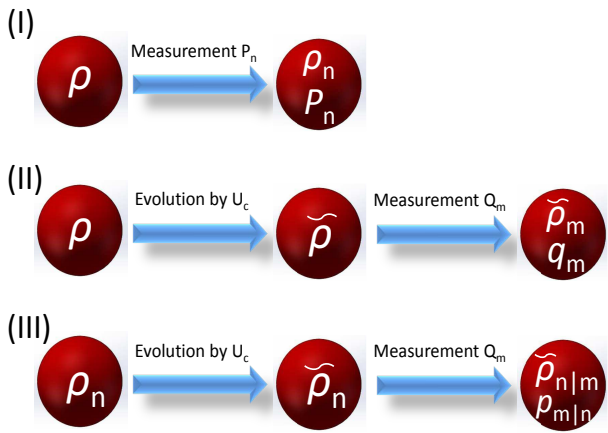}}
\caption{(Color online) (I) The first measurement. Measurement $P_{n}$ on the state $\rho$ to obtain $p_n$; (II) The second measurement. Measurement pulse under the operator $U_{C}$ yielding $\tilde{\rho}= U_{C}\rho U_{C}^{\dagger}$, along with a measurement $Q_{m}$ to obtain $q_m$; (III) Process for conditional probability. Measurement $Q_{m}$ on the state $\rho_n$ (produced from (I)), conditional on a previous measurement $P_{n}$, to obtain the conditional probability $p_{m|n}$. }
\label{Fig1}
\end{figure}

\begin{table}
\caption{Values for the measurement pulses implementing $P_{\pm}$ and $Q_{\pm}$ in the first part of the experiment with pure states, where $P=\sigma_z$ and $Q=\sigma_y$.}
\centering
\begin{tabular}{cccccccccccccccccccccccccccc}
\hline
\hline
  & ~~~$P_{+}$ & ~~~$P_{-}$ & ~~~$Q_{+}$ & ~~~$Q_{-}$ \\
  \hline
$\theta_2$ & ~~~0 &~~~$\pi$& ~~~ $\pi/2$ & ~~~$\pi/2$   \\

$\phi_2$ &  ~~~0 &  $~~~0$& ~~~0& ~~~$\pi$ \\
\hline
\hline
\end{tabular}
\label{Table1}
\end{table}

We first review briefly the main points in \cite{vedral}. The scheme gets started from a quantum state $\rho$, followed by a measurement on the basis $\{P\}$. Then the ensuing evolution is governed
by the most general completely positive trace preserving (CPTP) map, $\sum_{i}\Lambda_{i}(\cdot)\Lambda_{i}^{\dagger}$, followed by another measurement on the basis $\{Q\}$. Such a process, under the Born rule, can be described by the joint probability
\begin{equation}
p_{nm}=tr\{Q_{m}\sum_{i}\Lambda_{i}(P_{n}\rho P_{n})\Lambda_{i}^{\dagger}Q_{m}\}=p_{m|n}p_{n},
\label{eq1}
\end{equation}
where $p_{n}=tr\{P_{n}\rho\}$ is the probability regarding the measurement $\{P\}$, and $p_{m|n}=tr\{Q_{m}\sum_{i}(\Lambda_{i}P_{n}\Lambda_{i}^{\dagger})\}$ is the conditional probability implying the result of the second measurement dependent on the first measurement outcome. These quantities are associated with the mutual information
\begin{equation}
I_{nm}=-\ln q_{m} + \ln p_{m|n},
\label{eq2}
\end{equation}
which witnesses the difference between the entropy of the $m$th outcome without the knowledge of $n$ (given by -$\ln q_{m}$ with $q_m=tr\{Q_{m}\sum_i\Lambda_{i}\rho\Lambda_{i}^{\dagger}\}$) and the $m$th outcome when $n$ is known (given by -$\ln p_{m|n}$). Based on the mutual information $I_{nm}$, 
an information-theoretic equality is proposed, which satisfies the equality below,
\begin{equation}
\langle e^{-I_{nm}} \rangle :=\sum_{nm} p_{nm} e^{-I_{nm}} =1.
\label{eq3}
\end{equation}
The equation not only gives a simple expression of the probability conservation, but also represents a relation to the Jarzynski equality \cite{jarzynski}, if the system is initially prepared as a Gibbs state. The relation is stated as
\begin{equation}
I_{nm}=-\beta(W-\Delta F),
\label{eq6}
\end{equation}
where $W$ represents the work the system performs between the initial and final states with the free energy difference $\Delta F$. The free energy is defined as $F=-\ln Z/\beta$ with the partition function $Z=\sum_ne^{-\beta E_n}$, where $\beta=1/k_BT$ is the temperature parameter with the Boltzmann constant $k_B$ and the temperature $T$, and $E_{n}$ is the eigenenergy under the measurement.

Before presenting our experimental observations, we introduce briefly our system involving a single $^{40}Ca^{+}$ ion confined stably in a linear Paul trap \cite{sa}, whose axial and radial frequencies are $\omega_z/2\pi=1.01$ MHz and $\omega_r/2\pi=1.2$ MHz, respectively. Under the magnetic field of 6 Gauss, we encode the qubit in $|4 ^{2}S_{1/2}, m_{J}=+1/2\rangle$ as $\mid\downarrow\rangle$ and in $|3 ^{2}D_{5/2}, m_{J}=+3/2\rangle$ as $\mid\uparrow\rangle$, where $m_{J}$ is the magnetic quantum number. Although our investigation below only focuses on this qubit, cooling the ion to be ultracold is still necessary because thermal phonons yield offsets of Rabi oscillation. As such, the Doppler cooling and the resolved sideband cooling are executed in order, which leads to the z-axis motional mode to be cooled down to the vibrational ground state with the final average phonon number $\bar{n}_{z}<$ 0.1.

\begin{figure*}[hbtp]
\centering {\includegraphics[width=16 cm, height=4.5 cm]{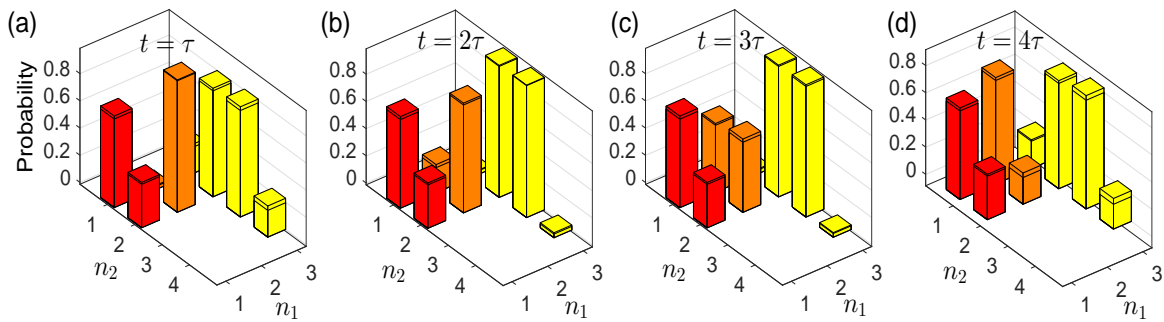}}
\caption{(Color online) Experimental results for the probabilities with pure states. In each panel, $n_1=1,n_2=1,2$ and $n_1=2,n_2=1,2$ denote the probabilities of $P_{-}$, $P_{+}$ and $Q_{-}$, $Q_{+}$, respectively, and $n_1=3,n_2=1,2,3,4$ correspond to the conditional probabilities $p_{-|-}$, $p_{-|+}$, $p_{+|-}$, $p_{+|+}$, respectively. The evolution time is set to $t=\tau,2\tau$,$3\tau$ and $4\tau$, with $\tau=\pi/5\Omega$. The initial state is $|\psi\rangle=(\sqrt{6}\mid\downarrow\rangle-i\sqrt{3}\mid\uparrow\rangle)/3$, and we obtain the data with the RMS error $\le$0.02 for individual point, under measurement repetition of 40,000 times. }
\label{Fig2}
\end{figure*}

\begin{table*}
\caption{Experimental values of the information-theoretic equality and the total mutual information using pure states, where $\sum_{nm}p_{nm}I_{nm}$ is to check whether the summation of all the possible mutual information is positive, and $\langle e^{-I_{nm}}\rangle$ should be close to unit. The numbers in parentheses represent the standard errors of the mean, i.e., the RMS error. }
\centering
\begin{tabular}{cccccccccccccccccccccccccccc}
\hline
\hline
\multirow{2}{*}{$\alpha$} & \multicolumn{4}{c}{$\sum_{nm}p_{nm}I_{nm}$}& &\multicolumn{4}{c}{$\langle e^{-I_{nm}}\rangle$}\cr
 \cline{2-5} \cline{7-10}
 & $t=\pi/5\Omega$ & $t=2\pi/5\Omega$ & $t=3\pi/5\Omega$& $t=4\pi/5\Omega$& &$t=\pi/5\Omega$ &$t=2\pi/5\Omega$ &$t=3\pi/5\Omega$&$t=4\pi/5\Omega$ \\
  \hline
1           &0.001(21)  &  0.002(6)  &   0.002(8)    &  0.001(16) & &   0.978(25) &  0.978(8)  &  0.978(11) & 0.973(20)  \\
$\sqrt{2/3}$&0.937(54)  &  0.560(23) &   0.508(19)   &  0.509(46) & &   0.985(39) &  0.985(61) &  1.015(63) & 0.974(29) \\
$\sqrt{1/3}$&0.520(36)  &  0.540(24) &   0.553(25)   &  0.930(51) & &   0.993(59) &  1.021(78) &  1.023(55) & 1.009(29) \\
\hline
\end{tabular}
\label{Table2}
\end{table*}

The qubit is initialized to $\mid\downarrow\rangle$ with a probability of $99.3(2)$\%. With the 729-nm laser pulses, we realize the carrier-transition Hamiltonian $H_c=\Omega(\sigma_+e^{i\phi}+\sigma_-e^{-i\phi})/2$ and the system evolves under the government of the carrier-transition operator
\begin{equation}
U_{C}(\theta,\phi) = \cos(\theta/2)I - i\sin(\theta/2)(\sigma_{x}\cos\phi -\sigma_{y}\sin\phi),
\label{eq4}
\end{equation}
where $\theta=\Omega t$ is determined by the evolution time with the laser-ion coupling strength $\Omega/2\pi=47.0(5)$ kHz, and $\phi$ represents the laser phase. Each experimental cycle is synchronized with the 50-Hz AC power line and repeated 40,000 times. The 729-nm laser beam is controlled by a double pass acousto-optic modulator.
The frequency sources for the acousto-optic modulator are based on a direct digital synthesizer controlled by a field programable gate array. Employment of the direct digital synthesizer helps the phase- and frequency-control of the 729-nm laser during each experimental operation.
\begin{table}
\caption{Values for the measurement pulses implementing $P_{\pm}$ and $Q_{\pm}$ in the second part of the experiment with Gibbs states, where $P_{\pm}$ and $Q_{\pm}$ are defined in the text, and $\mathcal{O}=(\sigma_x+\sqrt{3}\sigma_y)/2$.}
\centering
\begin{tabular}{cccccccccccccccccccccccccccc}
\hline
\hline
  & \multirow{2}{*}{$P_+$}& \multirow{2}{*}{$P_-$} & \multicolumn{3}{c}{$Q_+$}& &\multicolumn{3}{c}{$Q_-$}\cr
 \cline{4-6} \cline{8-10}
& & & $\sigma_x$ & $\sigma_y$ & $\mathcal{O}$& & $\sigma_x$ & $\sigma_y$ & $\mathcal{O}$ \\
  \hline
$\theta_2$ & 0  &$\pi$& $\pi/2$& $\pi/2$ & $\pi/2$& & $\pi/2$ &$\pi/2$&$\pi/2$   \\
$\phi_2$ &  0 &  $0$&$\pi/2$& 0& $\pi/6$& & $-\pi/2$&$\pi$ & $-5\pi/6$\\
\hline
\hline
\end{tabular}
\label{Table3}
\end{table}

\begin{table*}
\caption{Experimental results of the quantum Jarzynski equality and the total mutual information using Gibbs states. Here $H_f^i=E\sigma_x,E\sigma_y$ and $E(\sigma_x+\sqrt{3}\sigma_y)/2$ with $i=1,2,3$, respectively. We check whether $\sum_{nm}p_{nm}(\Delta F-W)$ is positive in the summation of all the possibilities, and $\langle e^{W-\Delta F}\rangle$ is close to unit. The numbers in parentheses represent the standard errors of the mean.}
\centering
\begin{tabular}{cccccccccccccccccccccccccccc}
\hline
\hline
\multirow{2}{*}{$\beta E$} & \multicolumn{3}{c}{$\sum_{nm}p_{nm}(\Delta F-W)$}& &\multicolumn{3}{c}{$\langle e^{W-\Delta F}\rangle$}\cr
 \cline{2-4} \cline{6-8}
 &$H_f^1$ & $H_f^2$  & $H_f^3$ &  &$H_f^1$ & $H_f^2$  & $H_f^3$\\
  \hline
0.2 &0.046(3)  &  0.044(4)  & 0.048(3)  & &   0.987(14) &  0.998(17) &  0.999(14)     \\
0.5 &0.234(8)  &  0.231(12) & 0.240(8)  & &   0.990(17) &  1.002(20) &  1.002(17) \\
1   &0.766(13) &  0.761(25) & 0.779(15) & &   0.963(23) &  0.977(26) &  0.976(24)    \\
\hline
\end{tabular}
\label{Table4}
\end{table*}

In the first part of our scheme, we focus on pure states to verify Eq. (\ref{eq3}) and the second part is to test the Jarzynski equality related to Eq. (\ref{eq6}) by exemplifying the thermal states as the Gibbs states. Our operations in each part consist of four steps \cite{sm}. For example, for the pure state case, the steps include: From $\mid\downarrow\rangle$ to $|\xi\rangle$ - state preparation; From $|\xi\rangle$ to $|\zeta\rangle$ - CPTP map; From $|\zeta\rangle$ to $|\varsigma\rangle$ - state measurement; Finally a projection measurement on $\mid\uparrow\rangle$. The first three steps are achieved, respectively, by $U_{C}(\theta_{0},\phi_{0})$, $U_{C}(\theta_{1},\phi_{1})$ and $U_{C}(\theta_{2},\phi_{2})$, based on Eq. (\ref{eq4}). The projectors, in the Bloch representation, are generally described as $P_{\pm}=(I\pm \vec{p}\cdot\vec{\sigma})/2$ and $Q_{\pm}=(I\pm \vec{q}\cdot\vec{\sigma})/2$ with $\vec{\sigma}=(\sigma_x,\sigma_y,\sigma_z)$.

For the case of pure states, we choose $\vec{p}=(0,0,1)$ and
$\vec{q}=(1,0,0)$. We first produce a pure state $\rho$ by $U_{C}(\theta_{0},\phi_{0})$, followed by a measurement $P_{\pm}=(I\pm\sigma_z)/2$, an ensuing evolution under $U_{C}(\theta_{1},\phi_{1})$ and another measurement $Q_{\pm}=(I\pm\sigma_y)/2$. Since our measurements are performed by detecting the population in the state $\mid\uparrow\rangle\langle\uparrow\mid$, execution of $P_{\pm}$ or $Q_{\pm}$ is accomplished by a measurement pulse under the unitary operator $U_{C}(\theta_{2},\phi_{2})$ in addition to a projective measurement. For example, the measurement pulse for $P_-$ is performed by $U_{C}^{\dagger}(\theta_{2},\phi_{2})\mid\uparrow\rangle\langle\uparrow\mid U_{C}(\theta_{2},\phi_{2})$ with $\theta_{2}=\pi,\phi_{2}=0$, as specified in Table \ref{Table1}. To verify Eq. (\ref{eq3}), we need three measurement results $p_n$, $p_m$ and $p_{m|n}$, which are obtained, respectively, by the three steps as shown in Fig. \ref{Fig1}.

With the pure state $\rho=|\psi\rangle\langle\psi|$ with $|\psi\rangle=\alpha\mid\downarrow\rangle-i\beta\mid\uparrow\rangle$ and $\alpha^2+\beta^2=1$,
we have accomplished experimental measurements $p_n$, $p_m$ and $p_{m|n}$ by choosing three different pure states with $\alpha=1$, $\sqrt{2/3}$ and $\sqrt{1/3}$. Fig. \ref{Fig2} demonstrates the results for $\alpha=\sqrt{2/3}$. In our case, since the first measurement is made on the eigenstates of $\sigma_{z}$, the results strongly depend on the initial state of the system and remain unchanged with time. But the second measurement is different due to outcomes from the eigenstates of $\sigma_{y}$.  As such, the results of both $Q_{\pm}$ and the conditional probability $p_{m|n}$ are time dependent. Based on the measurement results as listed in Table \ref{Table2}, we confirm Eq. (\ref{eq3}) under root-mean-square (RMS) error $\le 0.078$, in which the error is induced dominantly by quantum projection noise, relevant to vacuum fluctuation, rather than the thermal noise in conventional thermodynamics. This evidently indicates that Eq. (\ref{eq3}) is robust against vacuum fluctuation in quantum thermodynamical process. Besides, in terms of quantum information theory, the total mutual information should be never negative. But subject to quantum projection noise, individual observations of $I_{nm}$ in our experiment are sometimes negative \cite{sm}. Nevertheless, our observation of the total mutual information $\sum_{nm}p_{nm}I_{nm}$, as listed in Table \ref{Table2}, is always positive, in agreement with the results from the fluctuation theorem based on the probability distributions.

Considering a more general situation with the mixed states, we initially prepare a thermal state in the system, followed by a temporal evolution sandwiched by two projective measurements. In this way, we confirm a Jarzynski equality \cite{vedral,vedral1} relevant to the mutual information $I_{nm}$ as tested above. To this end, we may start from a thermal state $\rho_i=\exp(-\beta H_i)/Z_i$ with the partition function $Z_i=tr\{\exp(-\beta H_i)\}$, where $H_i$ is the Hamiltonian of the system after the projective measurement on $\{P\}$. So we assume $H_i=\sum_{\pm}E^i_{\pm}P_{\pm}$ with $\vert E^i_{\pm}\vert  =E^i$. For another projective measurement on $\{Q\}$, we have the Hamiltonian $H_f=\sum_{\pm}E^f_{\pm}Q_{\pm}$ with $\vert E^f_{\pm}\vert  =E^f$. In our experiment, due to only two levels involved, we simply have $E^{i}=E^{f}=E$. So the work is defined as $W=E_n^i-E_m^f$ where $E_n^i$ and $E_m^f$ are the corresponding eigenvalues regarding the measurements $\{P\}$ and $\{Q\}$. The free energy difference is $\Delta F=F_i-F_f$, where $F_k=-\ln Z_k/\beta$ with  $Z_k=tr\{\exp(-\beta H_k)\}$. Thus we have $p_{nm}=tr\{Q_m U_C P_n\rho_iP_nU_{C}^{\dagger}Q_m\}=tr\{Q_mP_n\rho_iP_n\} = tr\{Q_mP_n\rho_i\} = tr\{Q_mP_n\} e^{-\beta E_n^i}/Z_i$, where we have used the fact that $Q_m$ commutes with $U_{C}$ and $P_n$ commutes with $\rho_i$. Based on above processes, Eq. (\ref{eq6}) works and Eq. (\ref{eq3}) can be rewritten as \cite{vedral,vedral1}
\begin{equation}
\langle e^{\beta(W-\Delta F)}\rangle=1,
\label{Eq7}
\end{equation}
which is termed the Jarzynski equality to be verified below.

In our operations below, we choose $\vec{p}=(0,0,1)$, implying $H_i=E\sigma_z$, and we consider three different forms of $H_f$ with $\vec{q}=(1,0,0)$, $(0,1,0)$ and $(1/2,\sqrt{3}/2,0)$, respectively, corresponding to $H_f=E\sigma_x,E\sigma_y$ and $E(\sigma_x+\sqrt{3}\sigma_y)/2$. Then we obtain a two-level Gibbs state $\rho_i=\exp(-\beta E\sigma_z)/Z_i=[e^{\beta E}\mid\downarrow\rangle\langle\downarrow\mid+e^{-\beta E}\mid\uparrow\rangle\langle\uparrow\mid]/Z_i$ with $Z_i=e^{-\beta E}+e^{\beta E}$. In this case, we find that $Z_i=Z_f$ implying $\Delta F=0$.

By means of the qubit dephasing, we experimentally prepare the Gibbs state, and then carry out operations \cite{sm} following similar steps to the pure state case.
Accomplishment of the measurements regarding $P_{\pm}$ and $Q_{\pm}$ also depends on the measurement operator $U_C^{\dagger}(\theta_{2},\phi_{2})\mid\uparrow\rangle\langle\uparrow\mid U_{C}(\theta_{2},\phi_{2})$, where the values of $\theta_{2}$ and $\phi_{2}$ are listed in Table III. By considering the initial states regarding $\beta E=0.2, 0.5$ and $1.0$, respectively, we have carried out the above steps and confirmed Eq. (\ref{Eq7}) with high precision, see Table. \ref{Table4} where the RMS errors are smaller than $0.03$. Different from the case of pure states, both thermal noise and quantum projection noise exist in this case, where the latter is dominant as analyzed in \cite{sm}. We have a smaller RMS error here than the pure state case just because the measurements made here are simpler \cite{sm}. The observation values in Table. \ref{Table4} indicate that the Jarzynski equality holds under the influence of vacuum fluctuation and on the other hand, our operations are precise enough to witness a single-spin thermodynamic process governed by the Jarzynski equality.

The experimentally determined errors are partly from imperfection of initial state preparation (0.7(2)$\%$) and final state detection (0.22(8)$\%$). Decoherence effects are negligible due to our short-time implementation: 50 $\mu$s operation time for pure states and 3 ms operation time for mixed states. The dominant errors, as mentioned above, due to quantum projection noise are inevitable in any quantum mechanical measurement, but can be reduced by more measurements. As such, we have tried to repeat our measurements by 40,000 times, suppressing the relevant errors for individual point to be below 2$\%$.

In summary, our experiment has provided the first single-spin evidence confirming a simple and general equality involving the expectation value of the exponential of mutual information. Since the equality relies on the properties of classical probabilities (that arise from the projective quantum measurements) and is concomitant to the quantum Jarzynski equality, our experimental implementation at this fundamental level of a single spin will be helpful for further understanding thermodynamic processes in quantum regime, particularly when quantum information and more degrees of freedom are involved \cite {parrondo, vedral1}.

This work was supported by National Key R$\&$D Program of China under grant No. 2017YFA0304503, by National Natural Science Foundation of China under Grant Nos. 11734018, 11674360, 11404377, 91421111 and 11371247, and by the Strategic Priority Research Program of the Chinese Academy of Sciences under Grant No. XDB21010100. K.R. acknowledges thankfully support from CAS-TWAS president's fellowship. V.V. acknowledges funding from the National Research Foundation (Singapore), the Ministry of Education (Singapore), the Engineering and Physical Sciences Research
Council (UK), the Templeton Foundation and the Oxford Martin School. T.P.X. and L.L.Y. contributed equally to this work. \\

\end{document}